\documentclass[conference]{IEEEtran}

\ifCLASSINFOpdf
   \usepackage[pdftex]{graphicx}

\else
 
  \usepackage[dvips]{graphicx}
  
\fi

\usepackage{amsmath}
\usepackage{hyperref}

\usepackage[ruled]{algorithm2e}

\usepackage{array}
\author{\IEEEauthorblockN{Tamim Al Mahmud\IEEEauthorrefmark{1}, B M Mainul Hossain \IEEEauthorrefmark{2}, Dilshad Ara \IEEEauthorrefmark{3}}
\IEEEauthorblockA{Department of Computer Science and Engineering\IEEEauthorrefmark{1}\IEEEauthorrefmark{2}\\ 
Green University of Bangladesh, Dhaka-1207, Bangladesh\IEEEauthorrefmark{1}\\
University of Dhaka, Dhaka-1000, Bangladesh \IEEEauthorrefmark{2}\\
Dhaka International University, Dhaka-1205, Bangladesh\IEEEauthorrefmark{3} \\
Email: tamim@cse.green.edu.bd\IEEEauthorrefmark{1}, mainul@iit.du.ac.bd \IEEEauthorrefmark{2}, dilshadara1995@gmail.com\IEEEauthorrefmark{3}}}

\usepackage{eso-pic}
\newcommand\AtPageUpperMyright[1]{\AtPageUpperLeft{%
 \put(\LenToUnit{0.075\paperwidth},\LenToUnit{-1cm}){%
     \parbox{0.5\textwidth}{\raggedright\fontsize{9}{11}\selectfont #1}}%
 }}%
\newcommand{\conf}[1]{%
\AddToShipoutPictureBG*{%
\AtPageUpperMyright{#1}
}
}
\conf{
}

\begin{document}

\title{Automatic Reviewers Assignment to a Research Paper Based on Allied References and Publications Weight
}

\IEEEoverridecommandlockouts
\IEEEpubid{\makebox[\columnwidth]{Preprint version} \hspace{\columnsep}\makebox[\columnwidth]{ }}

\maketitle

\IEEEpubidadjcol

\begin{abstract}
Everyday there may be a vast stream of research documents give in to conferences, anthologies, journals, newsletters, annual report, daily paper, and different periodicals. Numerous such publications use unbiased and exterior specialists to survey. This methodology is frequently termed peer review, in addition the assessors are named referees. But it is not always possible to  pick  out the  best  referee for reviewing. Moreover, the new fields of research are emerging in every sector and the number of research is increasing dramatically. But, to review all of these papers, every journal assign a short team of referees and maybe they are not expert in all areas. For example, a research paper in the communication technology area should be reviewed by the expert from the same field. So, it is a big challenge to efficiently select the best reviewer or referee for a research paper.          
 
In this research work, we proposed and implemented a program or software where, we used a new strategy to select the best reviewers of a research work automatically. As we already know that every research paper contains references in the end of the paper and the references are generally from same area as the paper is. In this work, first of all, we collect the references and count the authors who have at least one paper that belong to the reference section. Once, we collect the author’s name, automatically browse the web to extract the research topic keywords. Then again search for finding top researchers in the specific topic field and count their h-index, i10-index and citations for first n author. After then, ranked top n author’s based on ranking score and automatically browse their homepage to retrieve emails address. We also check their co-authors and colleagues from the web and discard them accordingly from the list. The remaining tops n author’s (researcher) generally professor and maybe the best referees for reviewing a research paper. 

\textit{Automatic Frequency Finding Algorithm (AFFA), Automatic Citation Finding Algorithm (ACFA), Automatic Ranking Algorithm (ARA), and Automatic Email Finding Algorithm (AEFA)}

\end{abstract}

\IEEEpeerreviewmaketitle

\section{Introduction}
Publishing article is an essential part of a scholar’s proficient life. Though, lettering is not every one preferred activity, and accomplishment a paper issued can be a monotonous and slow process [1]. Fortunately, one important obstacle of selecting best reviewer (peer review) of the article make reproducing path can be easy by ensuing some diffident strategies and observes.  This research article offerings a combination of procedures in what way to find out a precise assessor. The article also summaries the procedure of issuing research papers in peer review journals and in the conference proceedings, targeting to deliver premature stage scholars with an accessible starter to essential complications of the assessor. The paper takings an interdisciplinary posture by a generous illustration from technology-enhanced knowledge research by implementing a program of several algorithms like AFFA (authors frequency finding algorithm), ACFA (authors citations finding algorithm), ARA (authors ranking algorithm) and AEFA (authors email finding algorithm).

\section{Related Work}\label{sec:related}
Peer review is an evaluation process for the competence, significance and originality of researches by qualified experts \cite{bts}. Many journals have an editorial team with an editor-in-chief and a number of scientific editors who are assigned responsibility for the peer review of individual papers \cite{hames2008peer}. These journals sometimes offer discussions before taking a reviewer. They just search the subject expert for selecting reviewers \cite{cooley1997web}. But their expert list is very small. Sometimes it is very hard to found a subject expert and it is time-consuming.  Generally, the techniques for distribution of article to a specific assessor can be characterized into two modules: preference-based approach and topic-based approach. 
\subsection{Preference Based Approach}

Numerous methods in the selection of assessor make it difficult to usage of favorite or stressed data from the assessors. In approach based on record preferences, the measures require the assessors to shot papers to understand whether they have their interest for the papers or not. Softness in this practice is the inefficiency of command statistics. Rigaux [5] suggested the use of collaborative filtering techniques to grow the preference by asking users    to bid on most of papers in a given topic. The basic hypothesis of collective intense procedures is that assessors who proposal likewise on a amount of the matching article have likely the similar predilection for supplementary papers.

\subsection{Topic Based Approach}
An interpretation of paper-to-assessor duties is that article must be allotted to assessors with a certain degree of knowledge in the subject of article. This understanding leader to topic-based practices that usages advantageous information. By using this data, commentator assignments can be finished in order to affirm a level of likeness between papers center and analyst’s investigation territory. The significant positioning of every commentator, grounded on topical information with esteem to a given paper, was entitled master finding or aptitude shaping. One reprobate stimulated with this strategy is to recognize what issues are canvassed in papers. Dumais and Nielsen \cite{dumais1992automating} matched papers to reviewers by Latent Semantic Indexing trained on reviewer-supplied abstracts. In Basu et al. \cite{basu2001technical}, abstracts from papers written by potential reviewers were extracted from the web via search engine, and then a Vector Space Model was constructed for the matching. Yarowsky and Florian \cite{yarowsky1999taking} extended this idea by a similar Vector Space Model with a Naïve Bayes Classifier. Wei and Croft \cite{wei2006lda} proposed a topic-based means by a language model with Dirichlet smoothing.

\section{METHODOLOGY }\label{sec:method}
The related works discussed about the reviewer selection that has been done manually by the editorial board. But, our aim is to do the work automatically by implement a software or program with some algorithms like AFFA, ACFA, and AEFA which will retrieve the correct person for reviewing the research paper. 

\subsection{Projected AFFA}
We need to extract authors name and frequency from reference section of a paper. But, PDF file is impossible to read by the compiler, whereas TEXT file is easily read by compiler. We were observing for a java grounded API to translate PDF to text, later working through several blogs; the PDFBox development came to our release. PDFBox is a lending library which can handle dissimilar types of PDF papers including converted PDF setups and extracts text and has a command line utility as well to convert PDF to text documents\cite{pdfbox}.  
\subsubsection{Logically split reference section for getting author’s name}  

As we want to find the author’s from the research paper references section and it was observed that each paper reference section separated by the keyword either capital case “REFERENCES” or in sentence cases “References”. Hence, we were trying to split the reference section by calling the function Split (“REFERENCES”) if there were two parts not found then try by Split (“References”). 
\subsubsection{Applying regular expression finding author’s name}

The regular expression helps us to split the references and provide the following format shown in figure 1. We observed that each reference contains author’s name, research title, journal name, volume, pages, year and also some carriage, newline and unnecessary values.

\begin{figure}[htb!]
	\centering
	\includegraphics[width=8cm,height=2.15cm]{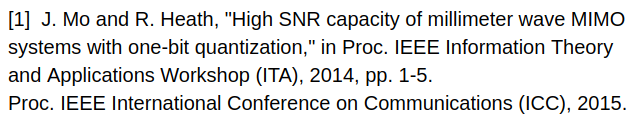}
	\caption{Reference Writing Format (Citation)}
	\label{fig:block}
\end{figure}

As, our target is to find the author’s from the paper which will not contain any carriage, newline, and numbers. So, it is mandatory to extract all the references in same format for understanding machine. Therefore, we applied some another regular expression that removed carriage, newline and all numbers and get references in the same structure.  
 
Now, all the references are in the same format, which is our expected format and if we see the structure of each reference (Figure 1) then we will observed that each author separated by a comma (,) and the paper title with start by double cote mark (“). Then, we called a new method that interfaces a hash-map, where String type for author’s name and Integer type for frequency. The HashMap class uses a hash-table to implement the Map interface \cite{dey2015program}. This allows the execution time of basic operations, such as get ( ) and put ( ), to remain constant even for large sets.  This procedure mapped all the parts split by the comma (,) of each reference.  
 
Then, all the parts of each reference containing commas are splits and stores into hash-map. After then, we deeply observed that many parts found that are separated by commas. Therefore, our challenged were to extract only the author’s name. Hence, to find out valid author’s, we introduced a method called checkStringValidity () \cite{dey2015program}. If the name starts with double cote (“) then we will delete it from the hash table. Because, we already know that the paper title starts with double cote (“) and we also delete keywords from the list if the name contains dictionary word that is not generally a name. 
 
\subsubsection{Counting Authors Frequency}  
After getting valid author’s list, then we can start author’s frequency counting from the hash-table. From the hash table when we found a new name we then put the value 1. If we found an author that is already in the hash table containing integer value we increased the value by one. This process will apply for the entire author’s in hash table. After assigning frequency value to each author then call the function entriesSortedByValues ( ) for sorting the authors based on frequency \cite{barua2016future}. Finally we get the result in format shown in figure 2.
\begin{figure}[htb!]
	\centering
	\includegraphics[width=8.8cm,height=4.2cm]{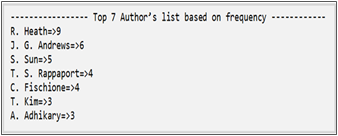}
	\caption{ List of Author’s with Frequency }
	\label{fig:block_2}
\end{figure}

\subsubsection{Projected ACFA} 
Now, our target is to find out selected author’s publications weight. To do we have tried to access information from Microsoft Academics, Research Gate and Google Scholar Account. After though a long observation we only able to access only the Google Scholar Account whereas the other two’s data are bound together in web server. After severely observation it was found that the Google Scholar Account was found by the following URLs only changed their parts of name by plus (+) operator shown in figure 3.\begin{figure}[htb!]
	\centering
	\includegraphics[width=8.8cm,height=3.5cm]{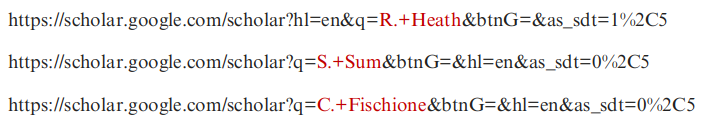}
	\caption{ Searching Format for Finding Selected Author’s Scholar Account  }
	\label{fig:block_3}
\end{figure}
If we clearly observe the figure 3 we can see that only changed in the portion where the author’s name append by a plus (+) operator. So, we can easily search an author just replacing that portion of the link with the required author name. After long scrutiny of the author’s Google Scholar Account source code, we found that every researcher has a unique id of 12 characters long, which is composed of the uppercase letter, lowercase letter, underscore, and numbers. Now, our challenged was to find the unique id for individual researcher or account holder. It was seen that the id number was found from the source code of the page after some logical splitting using regular expression shown in table 1.
 \begin{table}[!ht]
    \caption{Author’s Name with Google Scholar Account ID }
\setlength{\extrarowheight}{.5cm}
    \label{Tab:scholar}
    \centering
        \begin{tabular}{ | m{3cm}| m{4.5cm} |} 
        \hline
        Author's Name & ID (Google Scholar)\\ 
        \hline
        R. Heat & W\_ZpqUwAAAAJ\\ 
        \hline
        S. Sum & rrfl7UsAAAAJ\\ 
        \hline
        C. Fischione & RWGj7esAAAAJ\\
        \hline
        \end{tabular}
\end{table}

After getting the unique id, then we used a method named getPublicationInfo (Id). This method, redirect to the author’s Google Scholar account. Where, we found the Scholar name, University information, citations, h-index and i10-index. Then, our challenged were to retrieve all the information’s related to authors [13]. To do this we had split the source code into some logical points and also used regular expression. Finally, we get all the result in following format shown in table 2. 

\begin{table}[!ht]
    \caption{Author’s List with Extracting Publication Information’s from Web }
    \setlength{\extrarowheight}{.25cm}
    \label{Tab:scholar_1}
    \centering
    \begin{tabular}{ | m{1cm} | m{1cm} | m{1cm} | m{1cm} | m{1cm} | m{1cm} |} 
    \hline
    Name (Short) & Name & University & h-index & i10-index & Citations\\ 
    \hline
    R. Heat & James Robert Heath & California Institute of Technology & 99 & 220 & 54428\\ 
    \hline
    S. Sum & Shao-Cong Sun &Unknown Affiliation & 63 & 128 & 13624\\ 
    \hline
    C. Fischione & Carlo Fischione & Associate Professor, KTH Royal Institute of Technology & 23 & 48 & 1923\\
    \hline
    \end{tabular}
\end{table}

\subsubsection{Projected AEFA } 
Now, we have the full list of author’s along their publication weight but the main challenge is to extracting their emails from web. Some challenges include, it is very hard to find the home page or academic profile unless the author’s linked their profile in Google Scholar Account. Sometimes it’s not possible to find the right person from the sea of data in web. Some author binds the email and some are kept in image which is merely impossible to read by compiler. We know that an email address may contain the uppercase letter (A-Z), the lower case letter (a-z), numbers (0-9), and some special character like dot (.), underscore (\_ ), dash (-) and contains at or @ in the middle.  We were just trying to use this concept to solve all the challenges and overcome some of these. Finally capable of finding some author’s email by applying the code of RE (Regular Expression)

\section{EXPERIMENTAL RESULT ANAYSIS }
To experiment, we have taken three research article that are already published are listed below and the result (top three author’s) are shown table 3, 4, \& 5 respectively for data set 1, 2 \& 3.

\subsubsection{}A. Akusok, K. M. Bjork, Y. Miche and A. Lendasse, “High-Performance Extreme Learning Machines: A Complete Toolbox for Big Data Applications,”  IEEE Transaction, Published 30 June 2015, Volume 3, 2015 

\subsubsection{}B. Kehoe, S. Patil, P. Abbeel and K. Goldberg, “A Survey on Cloud Robotics and Automation,” IEEE transaction on Automation science and engineering, Vol. 12, No -2, April 20015 
 
\subsubsection{}K. Liu, L.Xu and J.Zhao, “Co-Extracting Opinion Targets and Opinion Words from Online Reviews Based on the Word Alignment Model” IEEE transaction on knowledge and data engineering, Vol 27, No-3, March 2015

\begin{table}[!ht]
    \caption{Final Output for Data Set 1}
        \setlength{\extrarowheight}{.5cm}
    \label{Tab:scholar_2}
    \centering
    \begin{tabular}{ | m{1cm} | m{0.5cm} | m{1.2cm} | m{2.5cm} | m{1.3cm} |} 
    \hline
    Author's (Rank by Score) & Total Score & Verified Email Domain & Homepage Link & Email\\ 
    \hline
    G. B. Huang & 1649 & ntu.edu.sg & \url{http://www.extreme-learning-machines.org} & \\ 
    \hline
    A. Lendasse & 408 & uiowa.edu & \url{http://www.engineering.uiowa.edu/mie/faculty-staff/amaury-lendasse} & \url{lendasse@uiowa.edu}\\ 
    \hline
    M. van Heeswijk & 44 & aalto.fi & \url{http://users. & .tkk.fi/heeswijk48} &\\
    \hline
    \end{tabular}
\end{table}

\begin{table}[!ht]
    \caption{Final Output for Data Set 2}
    \label{Tab:scholar_3}
        \setlength{\extrarowheight}{.5cm}
    \centering
    \begin{tabular}{ | m{1cm} | m{0.5cm} | m{1.2cm} | m{2.5cm} | m{1.3cm} |} 
    \hline
    Author's (Rank by Score) & Total Score & Verified Email Domain & Homepage Link & Email\\ 
    \hline
    K. Goldberg & 1116 & berkley.edu & \url{http://goldberg.berkeley.edu/} & \url{goldberg@berkeley.edu}\\ 
    \hline
    M. Beetz & 845 & in.tum.de & \url{http://ias.cs.tum.edu/people/beetz} & \url{lendasse@uiowa.edu}\\ 
    \hline
    R. D'Andrea & 637 & ethz.ch & \url{http://www.raffaello.name/} &\\
    \hline
    \end{tabular}
\end{table}

\begin{table}[!ht]
    \caption{Final Output for Data Set 3}
    \label{Tab:scholar_4}
        \setlength{\extrarowheight}{.5cm}
    \centering
    \begin{tabular}{ | m{1cm} | m{0.5cm} | m{1.2cm} | m{2.5cm} | m{1.3cm} |} 
    \hline
    Author's (Rank by Score) & Total Score & Verified Email Domain & Homepage Link & Email\\ 
    \hline
    H. Wang & 7682 & nshs.edu & \url{http://www.feinsteininstitute.org/Feinstein/About+Haichao+Wang} & \url{xwang8@sjtu.edu.cn}\\ 
    \hline
    K. Liu & 4545 & northwestern.edu & \url{http://www.tam.northwestern.edu/wkl/} & \\ 
    \hline
    Z. Liu & 3936 & tenorth.de & \url{http://www.tenorth.de/} &\\
    \hline
    \end{tabular}
\end{table}

\vspace{3mm}

\section{PERFORMANCE ANALYSIS}
The bar chart 1 shows the overall result that we found from our program for three experimental data. 

We see from the bar chart 1 [figure 4] and line chart [figure 5] that the accuracy of Frequency finding is 100 \%, ID 95\%, Email Domain 86\%, Professional Info 95\%, H-index 90\%, i10-index 90\%, citations 90\%, Homepage 71\%, Email 29\% and wrong Id selection only 10 \%. The overall accuracy is more that 80\%. So, we can easily say that it’s a great work for finding reviewers automatically.
 \begin{figure}[htb!]
	\centering
	\includegraphics[width=8.8cm,height=7.8cm]{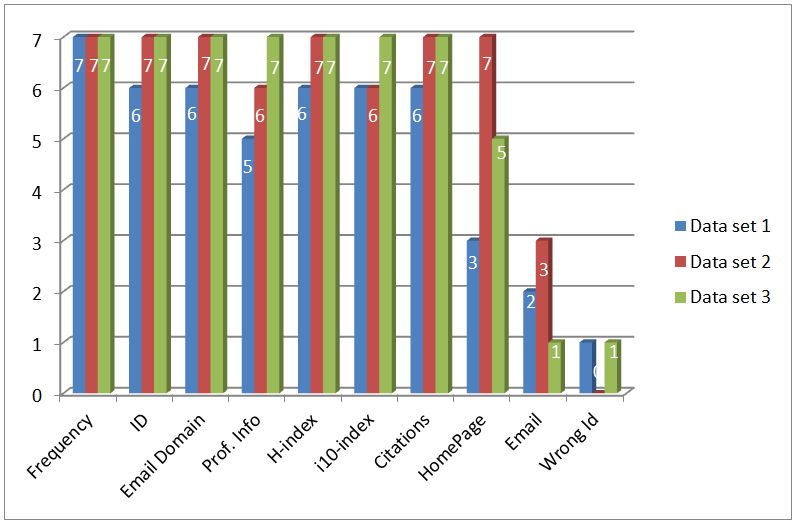}
	\caption{ Bar Chart 1:  Output Analysis of Data Set 1,2 \& 3}
	\label{fig:block_4}
\end{figure}
 \begin{figure}[htb!]
	\centering
	\includegraphics[width=8.8cm,height=8cm]{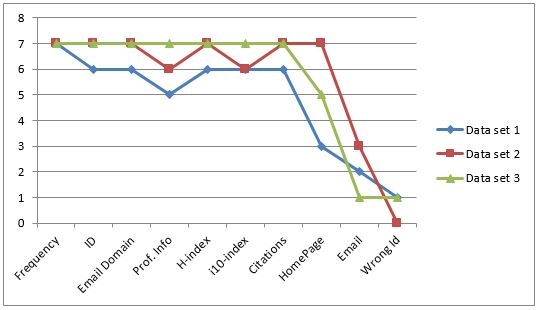}
	\caption{ Line Chart 1:  Output Analysis of Data Set 1,2 \& 3}
	\label{fig:block_5}
\end{figure}

Finally, if we can see from the figure of bar chart 2 that the accuracy of Frequency finding is 100\%, ID 95\%, Email Domain 86\%, Professional Info 95\%, H-index 90\%, i10-index 90\%, citations 90\%, Homepage 71\% and Email 29\%. Whereas wrong Id selection only 
10\% and the overall accuracy is more that 80\%. So, We can easily say that it’s a great work for finding reviewers automatically. The performance analysis also tells us about the number of correct values of each attribute.
 \begin{figure}[htb!]
	\centering
	\includegraphics[width=8.75cm,height=7.5cm]{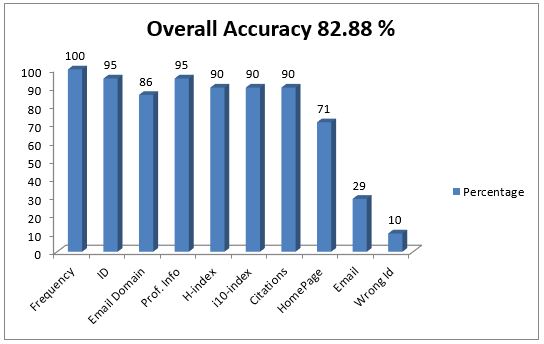}
	\caption{ Bar Chart 2:  Accuracy of Final Output of Data Set 1,2 \& 3}
	\label{fig:block_6}
\end{figure}

\section{Conclusion}
This thesis work is very much useful for every Journal like IEEE, ACM, Springer, and IGCA etc. to find reviewers automatically to review the research paper. Really it is a very tough job for chief editor to find the best reviewer for a research paper manually. To find the reviewer for a research paper the chief editor may follow a long manual process. 
\begin{itemize}
    \item Selected the paper keyword to know the field of the research area.
    \item Manually searching their reviewer list to find a professor in that field. 
    \item If not found in their list of the reviewer. Then search online for a professor in the same field of the paper. 
    \item Finally, select the reviewers. The process is time consuming. 
\end{itemize}
We just want to automate this process by applying some algorithm, where the chief editor of any research journal may use this program or software for finding best reviewers with emails automatically within a minute. Even, individual researcher can review their paper before going to publish. A concrete proof-of-concept implementation has been done perfectly to find the best reviewer for a research paper and validate by the performance analysis charts.

\bibliographystyle{IEEEtran}
\bibliography{references}

\end{document}